\documentclass[12pt]{article}
\usepackage{latexsym,amssymb,amstext}
\renewcommand{\baselinestretch}{1.15}

\newcommand{\be}{\begin{equation}}
\newcommand{\ee}{\end{equation}}
\newcommand{\beqs}{\begin{eqnarray}}
\newcommand{\eeqs}{\end{eqnarray}}

\def\({\left(}
\def\){\right)}

\def\mxth{\mathsurround=0pt }
\def\xversim#1#2{\lower2.pt\vbox{\baselineskip0pt \lineskip-.5pt
x  \ialign{$\mxth#1\hfil##\hfil$\crcr#2\crcr\sim\crcr}}}

\def\be{\begin{equation}}
\def\ee{\end{equation}}
\def\bea{\begin{eqnarray}}
\def\eea{\end{eqnarray}}

\newcommand{\ft}[2]{{\textstyle\frac{#1}{#2}}}

\def\bfone{\relax{\rm 1\kern-.35em 1}}

\makeatletter
\@addtoreset{equation}{section}
\makeatother

\begin{document}

\thispagestyle{empty}

\begin{flushright}\small
\end{flushright}

\bigskip

\begin{center}
  {\Large {\bf Actions for Non-Abelian Twisted Self-Duality}}
\end{center}


\vskip 8mm

\begin{center}
{\bf Henning Samtleben}\\[.4ex]
{\small Universit\'e de Lyon, Laboratoire de Physique, UMR 5672, CNRS et ENS de Lyon,\\
46 all\'ee d'Italie, F-69364 Lyon CEDEX 07, France \\[.5ex]
Institut Universitaire de France}
\end{center}

\vskip 8mm

\renewcommand{\baselinestretch}{1.05}
\begin{center} {\bf Abstract } \end{center}
{\small\begin{quotation}
The dynamics of abelian vector and antisymmetric tensor gauge fields can be described in
terms of twisted self-duality equations. These first-order equations relate the $p$-form 
fields to their dual forms by demanding that their respective field strengths are dual to each other.
It is well known that such equations can be integrated to a local action that carries on
equal footing the $p$-forms together with their duals and is manifestly duality invariant.
Space-time covariance is no longer manifest but still present with a non-standard 
realization of space-time diffeomorphisms on the gauge fields.
In this paper, we give a non-abelian generalization of this first-order action by gauging 
part of its global symmetries. The resulting field equations are non-abelian versions of
the twisted self-duality equations. A key element in the construction is the introduction
of proper couplings to higher-rank tensor fields.
We discuss possible applications (to Yang-Mills and supergravity theories)
and comment on the relation to previous no-go theorems.
\end{quotation}}
\renewcommand{\baselinestretch}{1.15}

\bigskip
\bigskip

\section{Introduction}

The dynamics of abelian vector and antisymmetric tensor gauge fields can be described in
terms of so-called twisted self-duality equations. A prominent example are Maxwell's equations: Rather
than expressing them in the standard form of second-order differential equations for a single gauge field, 
they may be written in terms of a gauge field and its magnetic dual by demanding that their
respective field strengths are dual to each other.
An analogous system of first-order equations can be formulated in general
for abelian $p$-forms in $D$-dimensionsal space-time and their dual $(D-p-2)$-forms.
It takes the schematic form
\bea
^*{\cal F} &=& \Omega\, {\cal M}\, {\cal F}
\;,
\label{TSD}
\eea
where ${\cal F}$ combines the abelian field strengths of the original $p$-forms and their magnetic duals,
and $\Omega\, {\cal M}$ denotes the `twist matrix' which squares to the same multiple 
of the identity as the Hodge star `\,$^*$\,' on the associated field strengths.
References~\cite{Cremmer:1997ct,Cremmer:1998px} have coined the term of `twisted self-duality equations'.
The system (\ref{TSD}) persists in generalized form in the presence of scalar fields, 
fermions, and Chern-Simons-type couplings as they typically arise in supersymmetric theories.
In particular, the twist matrix in the general case depends on the scalar fields of the theory.

The formulation of the $p$-form dynamics in terms of the first-order system (\ref{TSD})
plays an important role in exposing the duality symmetries of the theory. Of particular interest is the case
of `self-dual' $p$-forms in a $(2p+2)$-dimensional space-time.
In this case, the global duality symmetries typically mix the original $p$-forms
with their magnetic duals. Thus they remain symmetries of the equations of motion and 
only a subset thereof can be implemented in a local way as symmetries 
of the standard second-order action. A prime example is the electric/magnetic
duality of Maxwell's equations. Similarly, in maximal four-dimensional supergravity~\cite{Cremmer:1979up}, 
the 28 electric vector fields of the $N=8$ supermultiplet combine with their magnetic duals
into the fundamental 56-dimensional representation of the duality group $E_{7(7)}$,
of which only an $SL(8)$ subgroup is realized as a symmetry of the action of~\cite{Cremmer:1979up}.
For even $p$, the dynamics described by the system (\ref{TSD}) may not even be integrable to
a second-order action, as it happens for the chiral supergravities in 6 and 10 dimensions~\cite{Romans:1986er}, \cite{Schwarz:1983wa}.

It is known since the work of Henneaux-Teitelboim~\cite{Henneaux:1988gg}
and Schwarz-Sen~\cite{Schwarz:1993vs} that the twisted self-duality equations (\ref{TSD})
can be derived from a first-order action that carries on
equal footing the $p$-forms together with their dual fields and is manifestly duality invariant,
see~\cite{Deser:1976iy,Floreanini:1987as} for earlier work.
The price to pay for duality invariance is the abandonment of manifest general coordinate 
invariance of the action. Although not manifest, the latter may be restored with a non-standard 
realization of space-time diffeomorphisms on the gauge fields. 
Alternatively, these theories have been obtained as the gauge-fixed versions of non-polynomial 
Lagrangians with manifest space-time symmetry~\cite{Pasti:1995tn,Pasti:1996vs}.
Recent applications of such actions have led to a Lagrangian for $N=8$ supergravity
that possesses full $E_{7(7)}$ invariance, allowing to examine the role of 
the full $E_{7(7)}$ in perturbative quantization
of the theory~\cite{Hillmann:2009zf,Bossard:2010dq}.
Moreover, it has been advocated recently~\cite{Bunster:2011aw,Bunster:2011qp}
that the existence of such duality symmetric actions underlying (\ref{TSD}) may suggest to 
grant a more prominent role to duality covariance than to space-time covariance.
This in turn may go along with
the realization of the hidden symmetries underlying maximal 
supergravity theories~\cite{West:2001as,Damour:2002cu}.

In this paper, we present a non-abelian version of the twisted self-duality
equations (\ref{TSD}) and a first-order action from which it can be derived.
The action is constructed as a deformation of the abelian actions of 
\cite{Henneaux:1988gg,Schwarz:1993vs} by gauging part of their global symmetries.
We use the embedding tensor formalism of~\cite{Nicolai:2000sc,deWit:2002vt,deWit:2005ub},
which describes the possible gaugings in terms of a constant embedding tensor,
subject to a number of algebraic consistency constraints.
As a result, we find that the consistent gaugings of the first-order action
are constrained by precisely the same set of algebraic constraints as the gaugings
in the standard second-order formulation.
A key part in the construction is the introduction of couplings to
higher-order $p$-forms and a new topological term in the gauged theory.
In general, the global duality group is broken by the explicit choice of
an embedding tensor, but acts as a symmetry within the full class of gaugings.
I.e.\ the non-abelian Lagrangian remains formally invariant under the action of the
duality group, if the embedding tensor is treated as a spurionic object transforming
under this group. 

For transparency, we restrict the discussion in this paper to the case
of vector fields in four-dimensional space-time; in the last section we comment on
the (straightforward) generalization to $p$-forms in $D$ dimensions.
The construction of a first-order action for non-abelian gauge fields 
in particular allows to revisit the recent discussion of possible gaugings of (subgroups of) 
electric/magnetic duality~\cite{Bunster:2010wv,Deser:2010it}.
\bigskip

The paper is organized as follows: in sections~\ref{sec:twisted} and \ref{sec:actionabelian}, 
we briefly review the general structure of the abelian twisted self-duality equations for 
vector gauge fields in four space-time dimensions, and the associated duality-invariant
first-order action of~\cite{Henneaux:1988gg,Schwarz:1993vs}.
In section~\ref{sec:nonabelian}, we covariantize the abelian action 
upon gauging a subset of its global symmetry generators. 
The gauge group is characterized by a constant embedding tensor,
subject to a set of algebraic constraints. A key part in the construction of the covariant action
is the introduction of antisymmetric two-form tensor fields which turn out
to be the on-shell duals of the scalar fields of the theory.
The non-abelian action furthermore
comprises a topological term, introduced in section~\ref{sec:topological}.
In section~\ref{sec:action}, we present the full first-order action
and show that it gives rise to a non-abelian version of the
twisted self-duality equations~(\ref{TSD}), part of which now relate the scalar fields
and the two-form potentials. We furthermore show that although not manifest, the action
is invariant under four-dimensional coordinate transformations, if the action of space-time
diffeomorphisms on the vector and tensor 
gauge fields is properly modified by on-shell vanishing contributions.
In section~\ref{sec:YM} we illustrate the construction by a simple example:
pure Yang-Mills theory in absence of other matter couplings. 
The resulting first-order Lagrangian describes the Yang-Mills vector fields 
together with their duals. It gives rise to the Yang-Mills field equations and 
in the limit of vanishing gauge coupling constant consistently reduces to the
abelian action of~\cite{Henneaux:1988gg,Schwarz:1993vs}. Electric/magnetic duality
is explicitly broken by the choice of the embedding tensor. 
We close in section~\ref{sec:conclusions} with a discussion of possible applications
of these theories and comment on the relation to previous no-go theorems.

\section{Twisted self-duality equations}
\label{sec:twisted}

We start from the general second-order Lagrangian for 
$n$ `electric' vector fields $A_\mu{}^\Lambda$ with abelian field strengths 
in four space-time dimensions
${\cal F}_{\mu\nu}{}^\Lambda=2\partial_{[\mu}A_{\nu]}{}^\Lambda$
\begin{eqnarray}
  \label{L0}
\mathcal{L}_{0} &=&
 \ft14 \,e_4\, {\cal I}_{\Lambda\Sigma}\,\mathcal{F}_{\mu\nu}{}^{\Lambda} 
\mathcal{F}^{\mu\nu\,\Sigma} 
+\ft 18 {\cal R}_{\Lambda\Sigma}\;\varepsilon^{\mu\nu\rho\sigma} 
\mathcal{F}_{\mu\nu}{}^{\Lambda} 
\mathcal{F}_{\rho\sigma}{}^{\Sigma}  \;. 
\end{eqnarray}
The (possibly scalar dependent) symmetric matrices
$\mathcal{R}$ and $\mathcal{I}$ play the role of
generalized theta angles and coupling constants, respectively. 
We use a metric with signature $(-,+,+,+)$, $e_4 \equiv \sqrt{{\rm det}\,g}$ denotes the
square root of its determinant, and we use the completely antisymmetric 
Levi-Civita tensor density $\varepsilon_{0123}=1$.
The Bianchi identities and the equations of motion derived from (\ref{L0}) take the form
\begin{equation}
  \label{eom-Bianchi}
  \partial_{[\mu} \mathcal{F}_{\nu\rho]}{}^\Lambda  
  ~ =~ 0 ~=~ \partial_{[\mu} \mathcal{G}_{\nu\rho]\,\Lambda} \,, 
\end{equation}
respectively, 
where 
\begin{equation}
  \label{def-G}
  \mathcal{G}_{\mu\nu\,\Lambda} ~\equiv~ -  
  \varepsilon_{\mu\nu\rho\sigma}\,
  \frac{\partial \mathcal{L}_0}{\partial \mathcal{F}_{\rho\sigma}{}^\Lambda} 
  ~=~ {\cal R}_{\Lambda\Gamma} {\cal F}_{\mu\nu}{}^{\Gamma}
-\ft12e_4 \varepsilon_{\mu\nu\rho\sigma}\,
{\cal I}_{\Lambda\Gamma}\, {\cal F}{}^{\rho\sigma\,\Gamma}
  \;.
\end{equation}
The equations of motion thus take the form of integrability equations that 
(on a topologically trivial space-time that we shall assume in the following)
allow for the definition of the dual magnetic vector fields as 
${\cal F}_{\mu\nu\,\Lambda} \equiv
2\partial_{[\mu}A_{\nu]}{}_\Lambda \equiv {\cal G}_{\mu\nu\,\Lambda}$\,.
In order to set up an $\mathrm{Sp}(2n,\mathbb{R})$
covariant notation we introduce $2n$-dimensional symplectic indices
$M,N,\ldots$, such that $Z^M= (Z^\Lambda, Z_\Lambda)$.
The original equations of motion (\ref{eom-Bianchi}) can then be
rewritten in the form of covariant twisted self-duality equations~\cite{Cremmer:1979up,Cremmer:1997ct,Cremmer:1998px} 
for the symplectically covariant field strength 
${\cal F}^M \equiv ({\cal F}^\Lambda, {\cal F}_\Lambda)$\,:
\bea
{\cal F}_{\mu\nu}{}^M &=&
-\ft12\,e_4\varepsilon_{\mu\nu\rho\sigma}\,
\Omega^{MN}{\cal M}_{NK}(\phi^i)\,
{\cal F}^{\rho\sigma\,K}
\;,
\label{twisted}
\eea
with the positive definite symmetric matrix ${\cal M}_{MN}$
\bea
{\cal M}(\phi^i) &\equiv&
\left(
\begin{array}{cc}
-{\cal I}-{\cal R}{\cal I}^{-1}{\cal R}& 
{\cal R}{\cal I}^{-1}\\
{\cal I}^{-1}{\cal R} & -{\cal I}^{-1}
\end{array}
\right)
\;,
\label{defMeven}
\eea
and  the $\mathrm{Sp}(2n,\mathbb{R})$ invariant skew-symmetric tensor $\Omega_{MN}$
{}\footnote{
The conjugate matrix $\Omega^{MN}$ is defined by
$\Omega^{MN}\Omega_{NP}= - \delta^M{}_P$.
In the following, we use the symplectic matrix 
$\Omega^{MN}$ to raise and lower fundamental indices $M, N, \dots$
using north-west south-east conventions: $X^M=\Omega^{MN}X_N$, etc..}
\begin{equation}
  \label{eq:omega}
  \Omega = \left( \begin{array}{cc} 
0 & {\bf 1}\\ \!-{\bf 1} & 0 
\end{array} \right)  \;.  
\end{equation}
 By $\phi^i$ we denote the scalar fields of the theory
on which the matrix ${\cal M}_{MN}$ may depend.
Equation (\ref{twisted}) constitutes the explicit four-dimensional form 
of the general twisted self-duality equations~(\ref{TSD}).

The form of the twisted self-duality equations (\ref{twisted}) allows to  exhibit 
the properties of the global
duality symmetries of the theory \cite{Gaillard:1981rj} : 
Consider the generators~$t_\alpha$ of infinitesimal symmetries of the remaining matter 
couplings with linear action on the vector fields according to
\bea
\delta_\alpha  A_\mu{}^M &=& 
- (t_\alpha)_N{}^M\,A_\mu{}^N \;.
\label{varA}
\eea
The twisted self-duality equations (\ref{twisted}) are invariant under the action (\ref{varA})
if the action of $t_\alpha$ on the scalar fields of the theory induces a linear transformation
\bea
\delta_\alpha {\cal M}_{MN}(\phi^i) &=&
\delta_\alpha \phi^i \,\partial_i\, {\cal M}_{MN}(\phi) 
  ~\stackrel{!}{=}~
2(t_\alpha)_{(M}{}^K\,{\cal M}{}_{N)K}(\phi)\;,
\label{varM}
\eea
of the matrix ${\cal M}_{MN}(\phi)$ and moreover leaves the matrix $\Omega_{MN}$ invariant:
\bea
(t_\alpha)_M{}^K \Omega_{KN} &=& (t_\alpha)_N{}^K \Omega_{KM}
\;.
\label{symplectic}
\eea
The latter condition implies that the symmetry group must be embedded into the symplectic group:
$G_{\rm duality}\subset Sp(2n,\mathbb{R})$\,.
In the absence of scalar fields and thus for a constant matrix 
${\cal M}_{MN}$,
it follows directly from (\ref{varM}) that the duality symmetries must further be 
embedded into the compact subgroup $U(n)\subset Sp(2n,\mathbb{R})$.
For later use we denote the algebra of generators $t_\alpha$ of $G_{\rm duality}$ as
\bea
[t_\alpha,t_\beta]&=&f_{\alpha\beta}{}^\gamma\,t_\gamma
\;,
\label{algebra}
\eea
with structure constants $f_{\alpha\beta}{}^\gamma$.
Only the subset of triangular generators
\bea
(t_\alpha)_M{}^N &=& \left(
\begin{array}{cc}
*&*\\
0&*
\end{array}
\right)
\;,
\eea
is realized in a local way as symmetries of the second-order action~(\ref{L0}).
The remaining generators can be realized in a non-local way \cite{Deser:1976iy,Bunster:2011aw}.
A manifest off-shell realization of the full duality group $G_{\rm duality}$
is achieved by passing to an equivalent first-order action, reviewed in the next section.

\section{Action for abelian twisted self-duality}
\label{sec:actionabelian}

We first briefly review the first-order action for the abelian twisted
self-duality equations (\ref{twisted}) as originally constructed 
in~\cite{Henneaux:1988gg,Schwarz:1993vs} 
and recently revisited in \cite{Bunster:2011aw,Bunster:2011qp}. 
In the notation we closely follow reference~\cite{Hillmann:2009zf}.
As a first step, we split the four-dimensional coordinates into
$\{x^\mu\} \rightarrow \{x^0, x^i\}$ and the metric as
\bea
g_{\mu\nu} &=&  \left(
\begin{array}{cc}
-N^2+h_{ij} N^i N^j & h_{ij} N^i \\
h_{ij} N^j & h_{ij}
\end{array}
\right)
\;,
\nonumber
\eea
into the standard lapse and shift functions.
The twisted self-duality equations (\ref{twisted}) take the form
\bea
{\cal F}_{0i}{}^M - N^j {\cal F}_{ji}{}^M &=&
-\ft12 e_3\,\varepsilon_{ijk}\,N\,
\Omega^{MN}{\cal M}_{NK}\,
{\cal F}^{jk}{}^{K}
\;,
\label{twisted0}
\eea
where all spatial indices are raised with the metric $h^{ij}$\,.
With the definitions
\bea
{\cal E}_i{}^M &\equiv &{\cal F}_{0i}{}^M - N^j {\cal F}_{ji}{}^M
\;,\nonumber\\
{\cal B}_i{}^M &\equiv&-\ft12 e_3\,\varepsilon_{ijk}\,N\,
\Omega^{MN}{\cal M}_{NK}\,
{\cal F}^{jk}{}^{K}
\;,
\label{EB}
\eea
equations (\ref{twisted}) thus reduce to
\bea
{\cal E}_i{}^M &=& {\cal B}_i{}^M \;.
\label{eomEB}
\eea

These first-order equations are obtained from variation of the action
\bea
 {\cal L}_{\rm kin,ab} &=& \frac{e_3}{2N}\, h^{ij} \left( {\cal E}_i{}^M - {\cal B}_i{}^M\right) {\cal M}_{MN} \,{\cal B}_j{}^N
\;,
\label{actionHT}
\eea
where $e_3 = \sqrt{{\rm det}\,h}$ denotes the determinant of the three-dimensional vielbein. 
This action breaks the four-dimensional general coordinate covariance, 
but is manifestly invariant under abelian gauge transformations and 
under the action of the global duality 
symmetries (\ref{varA}), (\ref{varM}).
Variation of the action (\ref{actionHT}) gives rise to
\bea
\delta  {\cal L}_{\rm kin,ab} &=& 
\Omega_{MN} \, \varepsilon^{imn}\,
\partial_m \left({\cal E}_n{}^N - {\cal B}_n{}^N\right) \delta A_i{}^M
+\ft34 \,\Omega_{MN} \, \varepsilon^{imn}\,\partial_{[0} {\cal F}_{mn]}{}^N\,\delta A_i{}^M
\nonumber\\[.5ex]
&&{}
-\ft14 \,\Omega_{MN} \, \varepsilon^{imn}\,\partial_i {\cal F}_{mn}{}^N\,\delta A_0{}^M
+{\rm total~derivatives}
\;.
\label{deltaL0}
\eea
Upon taking into account the Bianchi identities, 
this shows that the Lagrangian (\ref{actionHT}) depends on the components $A_0{}^M$
only via a total derivative such that the manifestly gauge invariant action (\ref{actionHT})
can be replaced by an equivalent Lagrangian that only depends on the spatial
components $A_i{}^M$ of the vector fields. The action thus gives rise to the equations of motion
\bea
\varepsilon^{imn}\,
\partial_m \left({\cal E}^{\#}_n{}^N - {\cal B}_n{}^N\right) &=& 0\;,
\label{eomHT}
\eea
where ${\cal E}^{\#}_n{}^N\equiv {\cal E}_n{}^N+\partial_n A_0{}^M$
denotes the $A_0{}^M$-independent part of ${\cal E}_n{}^N$\,.
Integration of (\ref{eomHT}) on a topologically trivial space-time
then allows to define a function $A_0{}^M$ such that the original equations
of motion (\ref{eomEB}) are satisfied.
Moreover, it may be shown \cite{Henneaux:1988gg,Schwarz:1993vs,Bunster:2011qp}
that although not manifest,
the action (\ref{actionHT}) is invariant under four-dimensional coordinate reparametrization.

\section{\mbox{Non-abelian gauge theory and two-form potentials}}
\label{sec:nonabelian}

In this section, we start generalizing the above construction to the non-abelian case.
More precisely, we study the situation when a subgroup of the global symmetry group $G_{\rm duality}$
of the above theory is gauged. We use the
embedding tensor formalism~\cite{Nicolai:2000sc,deWit:2002vt,deWit:2005ub},
in which the gauging is parametrized by a constant embedding tensor $\Theta_M{}^\alpha$,
such that covariant derivatives are given by
\bea
D_\mu &\equiv& \partial_\mu -g {A}_\mu{}^M X_M ~\equiv~
 \partial_\mu - g{A}_\mu{}^M\, \Theta_M{}^\alpha \,t_{\alpha}
\;,
\label{covder}
\eea
with coupling constant $g$ and the gauge group generators $X_M\equiv\Theta_M{}^\alpha \,t_{\alpha}$
defined as linear combinations of the generators of the global duality group (\ref{algebra}).
In four space-time dimensions, the embedding tensor $\Theta_M{}^\alpha$ is constrained 
by the linear relations \cite{deWit:2005ub}
\bea
X_{(MNK)} &=& 0
\;,
\label{linear}
\eea
where we have defined the ``generalized structure constants"
\bea
X_{MN}{}^{K} &\equiv& (X_M)_N{}^K ~=~ \Theta_M{}^\alpha\,(t_\alpha)_N{}^K\;,
\label{gsc}
\eea
as the gauge group generators evaluated in the vector field representation.\footnote{
Note that unless they identically vanish, 
these generalized structure constants are never antisymmetric in the
first two indices, but satisfy $X_{(MN)}{}^K=\frac12\Omega^{KL}\Omega_{NP}X_{LM}{}^P$
due to the symplectic embedding~(\ref{symplectic}) and the linear constraint~(\ref{linear}).\label{fn:sym}}
Typically, the constraint (\ref{linear}) can be solved by projecting out 
some of the irreducible components of the tensor $\Theta_M{}^\alpha$ with respect to 
the global symmetry group $G_{\rm duality}$. It is therefore referred to as a `representation constraint'.\footnote{
In supersymmetric theories, this constraint typically also ensures supersymmetry of
the Lagrangian. In the $N=1$ theories, a non-vanishing component $X_{(MNK)}$
is related to the anomaly structure of the theory~\cite{DeRydt:2008hw}.}
Furthermore, the matrix $\Theta_M{}^\alpha$ is subject to the bilinear constraints
\begin{eqnarray}
  f_{\alpha\beta}{}^{\gamma}\, \Theta_{M}{}^{\alpha}\,\Theta_{N}{}^{\beta}
+(t_{\alpha})_{N}{}^{P}\,\Theta_{M}{}^{\alpha}\Theta_{P}{}^{\gamma} &=&0\,, 
  \label{eq:clos}  \\[1ex]
\Omega^{MN}\,\Theta_{M}{}^{\alpha}\Theta_{N}{}^{\beta}&=&0
\;,
\label{quadcon}
\end{eqnarray}
of which the first corresponds to a generalized Jacobi identity
and the second one insures locality of the gauging
(i.e.\ the existence of a symplectic frame in which all magnetic charges vanish).

The proper definition of covariant field strengths requires the
introduction of two-form potentials
$B_{\mu\nu\,\alpha}$~\cite{deWit:2004nw,deWit:2005hv,deWit:2005ub}
transforming in the adjoint representation of the
global duality group $G_{\rm duality}$. 
Explicitly, these field strengths are given by
\bea
{\cal H}_{\mu\nu}{}^M &\equiv&
 2\partial_{[\mu} {A}_{\nu]}{}^{M}
  + g(X_{N})_{P}{}^{M}
  \,{A}_{[\mu}{}^{N} {A}_{\nu]}{}^{P}
 -\ft12g \Theta^{M\alpha}\,B_{\mu\nu\,\alpha}
\;,
\label{defH}
\eea
with a St\"uckelberg-type coupling to the two-form potentials $B_{\mu\nu\,\alpha}$.
Later on, the field equations will identify these two-form potentials as the duals of the 
scalar fields of the theory, c.f.\ equation~(\ref{eomNA2}) below. 
Their introduction thus does not correspond to adding new degrees of freedom.
Local gauge transformations are given by
\bea
\delta_\Lambda {A}_\mu{}^M &=& D_\mu \Lambda^M   +\ft12 g \Theta^{M\alpha}\,\Lambda_{\mu\,\alpha}
\;,
\nonumber\\[1ex]
\delta_\Lambda B_{\mu\nu\,\alpha} &=&2D_{[\mu} \Lambda_{\nu]\alpha}
-2(t_\alpha)_{MN}\,\left(\Lambda^M {\cal H}_{\mu\nu}{}^N
- {A}_{[\mu}{}^M\, \delta_\Lambda {A}_{\nu]}{}^N\right)
\;,
\label{gaugeAB}
\eea
with parameters $\Lambda^M$ and $\Lambda_{\mu\,\alpha}$,
under which the field strengths (\ref{defH}) transform covariantly as
\bea
\delta_\Lambda {\cal H}_{\mu\nu}{}^M &=& -g \Lambda^K X_{KN}{}^M\,{\cal H}_{\mu\nu}{}^N
\;.
\label{varH}
\eea
The presence of the two-form fields in (\ref{defH}) is crucial for the covariant 
transformation behavior.
We finally mention that the covariant field strengths (\ref{defH}) satisfy the
generalized Bianchi identities
\bea
{D}_{[\mu}{\cal H}_{\nu\rho]}{}^M &=&
-\ft16g {\Theta}^{M\alpha}\,{\cal H}_{\mu\nu\rho\,\alpha}
\;,
\label{Bianchi2}
\eea
with the covariant non-abelian field strength ${\cal H}_{\mu\nu\rho\,\alpha}$
of the two-form tensor fields, given by
\begin{eqnarray}
{\cal H}_{\mu\nu\rho\,\alpha}&=&3 D_{[\mu}B_{\nu\rho]\alpha}
+6\,t_{\alpha PQ}\,A_{[\mu}{}^P\Big(\partial_\nu A_{\rho]}{}^Q
+\ft{1}{3}g X_{RS}{}^Q\,A_\nu{}^R A_{\rho]}{}^S\Big)+\dots\,.
\label{H3}
\end{eqnarray}
and satisfying in turn the Bianchi identities
\bea
{D}_{[\mu}{\cal H}_{\nu\rho\sigma]\alpha}{} 
&=& \ft32 (t_{\alpha})_{MN}\, {\cal H}_{[\mu\nu}{}^M {\cal H}_{\rho\sigma]}{}^N+\dots
\;.
\label{Bianchi3}
\eea
Here, the dots indicate terms that vanish after contraction with $\Theta_M{}^\alpha$
and thus remain invisible in this theory.

Covariantization of the action (\ref{actionHT}) with respect to the local gauge transformations
(\ref{gaugeAB}) is now straightforward: 
we redefine electric and magnetic fields by covariantizing the previous definitions (\ref{EB})
according to
\bea
{\cal E}_i{}^M &\equiv &{\cal H}_{0i}{}^M - N^j {\cal H}_{ji}{}^M
\;,\nonumber\\
{\cal B}_i{}^M &\equiv&-\ft12 e_3\,\varepsilon_{ijk}\,N\,
\Omega^{MN}{\cal M}_{NK}\,
{\cal H}^{jk}{}^{K}
\;,
\label{EBcov}
\eea
and define the kinetic Lagrangian by covariantizing (\ref{actionHT}) as
\bea
{\cal L}_{\rm kin} &=& 
\frac{e_3}{2N}\, h^{ij} \left( {\cal E}_i{}^M - {\cal B}_i{}^M\right) {\cal M}_{MN} \,{\cal B}_j{}^N
\;,
\label{kincov}
\eea 
with the new covariant ${\cal E}_i{}^M$, ${\cal B}_i{}^M$\,.
In this form, the kinetic term is manifestly invariant under the non-abelian gauge transformations
(\ref{gaugeAB}), since the fields ${\cal E}_i{}^M$ and ${\cal B}_i{}^M$ transform covariantly
according to (\ref{varH}) and the matrix ${\cal M}_{MN}$ transforms according to (\ref{varM}).
A short calculation shows that the general 
variation of the new covariant kinetic Lagrangian 
gives rise to
\bea
\delta  {\cal L}_{\rm kin} &=& 
\Omega_{MN} \, \varepsilon^{imn}\,
D_m \left({\cal E}_n{}^N - {\cal B}_n{}^N\right) \delta A_i{}^M
+\ft18 \, \varepsilon^{imn}\,g\Theta_M{}^\alpha\,{\cal H}_{0mn\,\alpha}
\,\delta A_i{}^M
\nonumber\\[.5ex]
&&{}
-\ft1{24} \,\varepsilon^{imn}\,
g\Theta_M{}^\alpha \,{\cal H}_{imn\,\alpha}\,\delta A_0{}^M
-\ft18\, \varepsilon^{imn}\,g\Theta_M{}^\alpha
\left({\cal E}_i{}^M - 2{\cal B}_i{}^M\right) \Delta B_{mn\,\alpha}
\nonumber\\[.5ex]
&&{}
+\ft18
\, \varepsilon^{imn}\,g\Theta_M{}^\alpha\,N^j {\cal H}_{ji}{}^M
 \Delta B_{mn\,\alpha}
+\ft18
\, \varepsilon^{imn}\,g\Theta_M{}^\alpha\, {\cal H}_{mn}{}^M
 \Delta B_{0i\,\alpha}
\;,
\label{deltaL}
\eea
up to total derivatives. Here, we have used the Bianchi identities (\ref{Bianchi2}),
and introduced the `covariant variations' $\Delta B_{\mu\nu\,\alpha}
\equiv \delta B_{\mu\nu\,\alpha} -2(t_{\alpha})_{MN}\,A_{[\mu}{}^M\delta A_{\nu]}{}^N$,
for a more compact notation.
The variation (\ref{deltaL}) cannot yet be the final answer.
E.g., as it stands, the field equations induced by this Lagrangian imply vanishing of the field strength
$\Theta_M{}^\alpha {\cal H}_{mn}{}^M$ in contrast to the desired dynamics.
Also variation w.r.t.\ $B_{mn\,\alpha}$ induces first-order field equations that contradict (\ref{eomEB}).
To cure these inconsistencies,
the theory needs to be amended by a topological term that we will give in the following.

\section{The non-abelian topological term}
\label{sec:topological}

Let us consider the following topological term in order of the gauge coupling constant
\bea
{\cal L}_{\rm top} &=&
\frac1{16}\,\varepsilon^{\mu\nu\rho\sigma} \,g\Theta_M{}^\alpha\,
 B_{\mu\nu\,\alpha}\, {\cal H}_{\rho\sigma}{}^M
 +\frac1{12}\,\varepsilon^{\mu\nu\rho\sigma} \,gX_{PQM}\,\partial_\mu A_{\nu}{}^M A_{\rho}{}^P A_{\sigma}{}^Q
\;,
\label{Ltop}
\eea
that is manifestly four-dimensional space-time covariant and does not depend on the space-time
metric nor on the scalar fields of the theory.
Similar terms have appeared in~\cite{deWit:2005ub} for general gaugings with magnetic charges and in
\cite{deWit:1984px} without the two-forms for gaugings of triangular subgroups of $G_{\rm duality}$.
Here, the structure of this term is considerably simpler due to the symplectic covariance 
of the formalism.\footnote{
Note in particular the absence of the 
$A^4$ and the $B^2$ terms in (\ref{Ltop}) and (\ref{varLtop}), which is
due to the identities (\ref{linear}), (\ref{quadcon}) and their consequences such as
\bea
2X_{MQ[P}\,X_{RS]}{}^Q\,+ 3X_{QM[P}\, X_{RS]}{}^{Q}  &=& 0 \;,
\nonumber
\eea
etc..}
In particular, this topological term is separately gauge invariant.
The general variation of (\ref{Ltop}) is given by
\bea
\delta {\cal L}_{\rm top} &=&
-\frac1{24}\varepsilon^{\mu\nu\rho\sigma} \,g\Theta_M{}^\alpha\,\left(
 {\cal H}_{\mu\nu\rho\,\alpha}\,\delta A_{\sigma}{}^M
-\ft32{\cal H}_{\mu\nu}{}^M \,\Delta B_{\rho\sigma\,\alpha} 
\right)
\;,
\label{varLtop}
\eea
from which one readily  deduces its invariance under local gauge transformations (\ref{gaugeAB})
according to
\bea
\delta_\Lambda {\cal L}_{\rm top} &=&
- \frac3{8}\,\varepsilon^{\mu\nu\rho\sigma} \,
g \Lambda^K\,X_{(KMN)}\, {\cal H}_{\mu\nu}{}^M {\cal H}_{\rho\sigma}{}^N   
~=~0\;,
\eea
with the Bianchi identities (\ref{Bianchi2}), (\ref{Bianchi3}) and the linear constraint (\ref{linear}).

Upon the $3+1$ split of the coordinates, the variation of the topological term 
nicely combines with the variation (\ref{deltaL}) of the kinetic term, and together 
they give rise to
\bea
\delta \left(
{\cal L}_{\rm kin}+{\cal L}_{\rm top}
\right)
&=& 
\varepsilon^{imn}\,\left\{
\Omega_{MN}D_m \left({\cal E}_n{}^N - {\cal B}_n{}^N\right)
+\ft14 g\Theta_M{}^\alpha {\cal H}_{0mn\,\alpha} \right\} \delta A_i{}^M
\nonumber\\[.5ex]
&&{}
-\ft1{12} \,\varepsilon^{imn}\,
g\Theta_M{}^\alpha \,{\cal H}_{imn\,\alpha}\,\delta A_0{}^M
\nonumber\\[.5ex]
&&{}
-\ft14\, \varepsilon^{imn}\,g\Theta_M{}^\alpha
\left({\cal E}_i{}^M - {\cal B}_i{}^M\right) \Delta B_{mn\,\alpha}
\;,
\label{varLL}
\eea
up to total derivatives. The relative factor between 
${\cal L}_{\rm kin}$ and ${\cal L}_{\rm top}$ is chosen such that the combined action 
does not depend on the components $B_{0i\,\alpha}$ while 
it depends on the components $B_{ij\,\alpha}$ and $A_0{}^M$ only upon projection with
the embedding tensor via $\Theta_M{}^\alpha B_{ij\,\alpha}$ and $\Theta_M{}^\alpha A_0{}^M$, respectively.
This is the proper generalization of the $A_0{}^M$-independent action (\ref{actionHT}) in the abelian case.
Let us also note, that in the full gauged theory, the field equations will receive further contributions 
from variation of the matter Lagrangian that is now coupled to the gauge fields:
\bea
\frac{\delta {\cal L}_{\rm matter}}{\delta A_\mu{}^M} &\equiv& e_4\,g\, j^\mu{}_M
\;,
\label{defJ}
\eea
with the covariant currents $j^\mu{}_M$\,.
We will show in the next section, that the field equations induced by the variation (\ref{varLL}), (\ref{defJ})
give the proper non-abelian extension of the twisted self-duality equations.

\section{Action and equations of motion}
\label{sec:action}

The full non-abelian Lagrangian thus is given by the sum of (\ref{kincov}), (\ref{Ltop})
and possible matter couplings (scalars, fermions, gravity)
\bea
{\cal L}_{\rm covariant}
&=&
\frac{e_3}{2N}\, h^{ij} \left( {\cal E}_i{}^M - {\cal B}_i{}^M\right) {\cal M}_{MN} \,{\cal B}_j{}^N
\nonumber\\[.5ex]
&&{}+
\frac1{16}\,\varepsilon^{\mu\nu\rho\sigma} \,g\Theta_M{}^\alpha\,
 B_{\mu\nu\,\alpha}\, {\cal H}_{\rho\sigma}{}^M
 +\frac1{12}\,\varepsilon^{\mu\nu\rho\sigma} \,gX_{PQM}\,\partial_\mu A_{\nu}{}^M A_{\rho}{}^P A_{\sigma}{}^Q
 \nonumber\\[1.5ex]
&&{}+
{\cal L}_{\rm matter}
\;,
\label{Lfull}
\eea
with the covariant electric and magnetic fields ${\cal E}_i{}^M$, ${\cal B}_i{}^M$ from (\ref{EBcov}).
The variation of (\ref{Lfull}) is given in (\ref{varLL}), (\ref{defJ}). 
Formally, the Lagrangian (\ref{Lfull}) is invariant under the full action of the global
duality group $G_{\rm duality}$, if the embedding tensor $\Theta_M{}^\alpha$  
is treated as a spurionic object transforming
under $G_{\rm duality}$. A concrete choice of the embedding tensor 
will specify a particular gauging and explicitly break $G_{\rm duality}$.
The choice of the embedding tensor is restricted by the 
algebraic consistency conditions (\ref{linear}), (\ref{quadcon}). 
These coincide with the constraints imposed on the embedding tensor in 
the second-order formulation of the theory~\cite{deWit:2005ub}.
We will show now that the action~(\ref{Lfull}) implies the following set of equations of motion
\bea
{\cal E}_i{}^M &=& {\cal B}_i{}^M
\;,\label{eomNA1}\\
\Theta_M{}^\alpha\,{\cal H}_{\mu\nu\rho\,\alpha} &=&
-2 e_4 \varepsilon_{\mu\nu\rho\sigma}\,j^\sigma{}_M\;,
\qquad
\mbox{with}\quad
j^\sigma{}_M
 ~\equiv~ e_4^{-1}g^{-1}\,\frac{\delta {\cal L}_{\rm matter}}{\delta A_\sigma{}^M}\;,
\label{eomNA2}
\eea
of which the first is the non-abelian version of the twisted self-duality equation (\ref{eomEB}),
and the second is the four-dimensional duality equation relating the two-form fields to the scalar fields or, more precisely,
to the Noether current of the invariances that have been gauged.
These equations parallel the equations of motion 
obtained from the second-order formalism with magnetic gauge fields of \cite{deWit:2005ub}.

In order to derive the equations of motion (\ref{eomNA1}), (\ref{eomNA2}) 
we proceed in analogy with the abelian case
of section~\ref{sec:actionabelian} and make use of the fact that the full Lagrangian does not depend 
on the components $B_{0i\,\alpha}$ and depends on the 
$A_0{}^M$ only upon projection with the embedding 
tensor as $\Theta_M{}^\alpha A_0{}^M$. We then show that the equations of motion for the remaining
fields precisely allow for the introduction of the missing $B_{0i\,\alpha}$ and $A_0{}^M$ such that 
equations (\ref{eomNA1}), (\ref{eomNA2}) are satisfied.
To this end, we first choose a basis of vector fields $A_\mu{}^M$ and of 
global symmetry generators $t_\alpha$, such that the embedding tensor $\Theta_M{}^\alpha$
of the theory takes the block diagonal form 
\bea
\Theta_M{}^\alpha &=& \left(
\begin{array}{cc}
\Theta_A{}^a & \Theta_A{}^{\underline{a}}\\
\Theta_{\underline{A}}{}^a & \Theta_{\underline{A}}{}^{\underline{a}}
\end{array}
\right)
~\equiv~
\left(
\begin{array}{cc}
\Theta_A{}^a &0\\
0 & 0
\end{array}
\right)
\;,
\eea
with invertible $\Theta_A{}^a$\,.
I.e.\ only the gauge fields $A_\mu{}^A$ participate in the gauging and they
gauge the subgroup spanned by generators $t_a$. The quadratic constraint (\ref{quadcon})
implies that in this basis $\Omega^{AB}=0$\,. 
In analogy to (\ref{eomHT}) we denote by ${\cal E}^{\#}_i{}^M$ 
the part of ${\cal E}_i{}^M$ which is independent of $B_{0i\,\alpha}$ and 
$A_0{}^{\underline{A}}$\,. We note that in the above basis ${\cal E}_i{}^A={\cal E}^{\#}_i{}^A$\,.
Then, variation of (\ref{Lfull}) w.r.t.\ $B_{ij\,\alpha}$ according to (\ref{varLL})
gives rise to the equations 
\bea
{\cal E}_n{}^A &=& {\cal B}_n{}^A
\;,
\label{EBf1}
\eea
Next, from variation w.r.t.\ $A_i{}^{\underline{A}}$ we obtain
\bea
0~=~\epsilon^{imn}\,\Omega_{\underline{A}\underline{B}}\, D_m \left({\cal E}^{\#}_n{}^{\underline{B}} - {\cal B}_n{}^{\underline{B}}\right)
 &=& 
 \epsilon^{imn}\,\Omega_{\underline{A}\underline{B}} 
 \,\partial_m \left({\cal E}^{\#}_n{}^{\underline{B}} - {\cal B}_n{}^{\underline{B}}\right)
 \;,
\eea
where the second equality makes use of (\ref{linear}) and (\ref{EBf1}).
From this equation we deduce (in analogy to (\ref{eomHT})) 
the existence of a function $A_0{}^{\underline{B}}$ such that
\bea
\Omega_{\underline{A}\underline{B}} \,
 \left({\cal E}^{\#}_n{}^{\underline{B}} - {\cal B}_n{}^{\underline{B}}\right) &\equiv&
 \Omega_{\underline{A}\underline{B}} \,\partial_n A_0{}^{\underline{B}}
 \qquad\Longrightarrow\qquad
 \Omega_{\underline{A}\underline{B}} \,
 {\cal E}_n{}^{\underline{B}} ~=~  \Omega_{\underline{A}\underline{B}}\,  {\cal B}_n{}^{\underline{B}}
 \;.
 \label{EBf2}
\eea
In general, the matrix $\Omega_{\underline{A}\underline{B}}$ is not invertible such 
that $A_0{}^{\underline{B}}$ is not uniquely defined by this equation. The ambiguity
precisely corresponds to the freedom of a gauge transformation (\ref{gaugeAB}) with 
parameter $\Lambda_{0\,a}$\,.
Finally, we may choose the component $B_{0i\,a}$ such that 
\bea
\Omega_{A\underline{A}}\,{\cal E}_n{}^{\underline{A}} &=&  
\Omega_{A\underline{A}}\,{\cal B}_n{}^{\underline{A}} 
\;,
\label{EBf3}
\eea
with the $A_0{}^{\underline{B}}$ chosen in (\ref{EBf2}).
Together, equations (\ref{EBf1}), (\ref{EBf2}), and (\ref{EBf3}), build the non-abelian
twisted self-duality equation (\ref{eomNA1}). 
The remaining parts of the $\delta A_0{}^A$ and $\delta A_i{}^A$ variations of (\ref{varLL}),
(\ref{defJ}) combine into the second duality equation (\ref{eomNA2}).
We have thus shown, that the field equations induced by the covariantized 
Lagrangian (\ref{Lfull})
give rise to the set of non-abelian field equations (\ref{eomNA1}), (\ref{eomNA2}).

Let us finally discuss the symmetries of the action (\ref{Lfull}).
Built as the sum of two gauge invariant terms, the action is clearly
invariant under the local gauge transformations (\ref{gaugeAB}).
Although four-dimensional coordinate invariance is not manifest,
it follows from (\ref{varLL}) after some calculation,
that the action is also invariant under time-like diffeomorphisms~$\xi^0$,
provided the gauge field potentials transform as
\bea
\delta A_i{}^M &=& \xi^0\,\left({\cal B}_i{}^M+N^j {\cal H}_{ji}{}^M\right)\;,
\nonumber\\
\delta A_0{}^M &=& 0\;,
\nonumber\\
\Theta_M{}^\alpha \Delta B_{ij\,\alpha} &=& -2 e_3 N \xi^0 \varepsilon_{ijk}\,j^k{}_M
\;.
\label{moddiff}
\eea
On-shell, i.e.\ upon using the equations of motion (\ref{eomNA1}), (\ref{eomNA2}),
these transformation laws reduce to
\bea
\delta A_i{}^M &\approx& \xi^0\,{\cal H}_{0i}\;,
\quad
\Theta_M{}^\alpha \delta B_{ij\,\alpha} ~\approx~ \Theta_M{}^\alpha\, \xi^0 {\cal H}_{0ij\,\alpha}
+2(t_\alpha)_{MN} A_{[i}{}^M \delta A_{j]}{}^N
\;,
\eea
which in turn reproduce the standard transformation behavior under time-like
diffeomorphisms up to local gauge transformations (\ref{gaugeAB})
with parameters $\Lambda^M \equiv -\xi^0 A_0{}^M$, $\Lambda_{\mu\,\alpha}\equiv
-B_{\mu i\,\alpha}-(t_\alpha)_{KL}A_\mu{}^KA_0{}^L $\,.
In the abelian case, the modified form of time-like diffeomorphisms (\ref{moddiff})
for the vector fields $A_i{}^M$ coincides with the
results of~\cite{Henneaux:1988gg,Schwarz:1993vs}.
The main difference in the non-abelian case (apart from standard gauge covariantization)
is the explicit appearance of the gauge fields in
the matter part ${\cal L}_{\rm matter}$ 
of the Lagrangian. Since these couplings are manifestly four-dimensional
coordinate covariant, the off-shell modification of the
transformation law (\ref{moddiff}) leads to extra contributions
from this sector which are proportional to the equations of motion
\bea
\delta_{\rm extra} \,{\cal L} &=& 
\frac{\delta {\cal L}_{\rm matter}}{\delta A_i{}^M}\,
\xi^0 \left( {\cal B}_i{}^M-{\cal E}_i{}^M \right)
\;.
\eea
With (\ref{varLL}) and (\ref{defJ}), these contributions are precisely cancelled 
by the modified transformation law of $B_{ij}{}^M$ in (\ref{moddiff}).
To summarize, the Lagrangian (\ref{Lfull}) is invariant under local 
gauge transformations (\ref{gaugeAB}) and also under four-dimensional diffeomorphisms
with the modified transformation laws (\ref{moddiff}).

\section{Example: Yang-Mills theory}
\label{sec:YM}

The construction of a first-order action for non-abelian gauge theories presented above
hinges on the solution of the algebraic consistency conditions (\ref{linear}), (\ref{quadcon}) for the 
embedding tensor $\Theta_M{}^\alpha$. We stress once more that these conditions
are precisely the same as in the second-order formulation of the theory~\cite{deWit:2005ub}.
In particular, the solutions of these equations encode
all possible gaugings of a given abelian theory. 
The most interesting examples of such theories arise from extended gauged 
supergravities~\cite{Schon:2006kz,deWit:2007mt}, for which solutions to (\ref{linear}), (\ref{quadcon})
can be found by group-theoretical methods.
E.g.\ the presented construction immediately gives rise to a first-order action
for the maximal $SO(8)$ gauged supergravity of~\cite{deWit:1982ig}, obtained
as a deformation of the abelian first-order action of~\cite{Hillmann:2009zf}.

In this section, as an illustration of the construction we will discuss a much simpler 
(and somewhat degenerate) example: pure Yang-Mills theory in the absence of further matter content.
The starting point for the construction of this theory is
the abelian action (\ref{L0}) for $n$ vector fields with constant ${\cal I}$, and ${\cal R}=0$,
corresponding to $n$ copies of the standard Maxwell action. 
Accordingly, the matrix ${\cal M}_{MN}$ is constant, such that
the global symmetry of the twisted self-duality equations (\ref{twisted}) and the first-order
action (\ref{actionHT}) is the compact $U(n)$. Its $U(1)$ subgroup is the standard electric/magnetic
duality. Possible gaugings of this theory are described by an embedding tensor according to 
(\ref{covder}) which is a $2n\times n^2$ matrix. A quick counting shows that the linear 
constraints (\ref{linear})
reduce the number of independent components in $\Theta_M{}^\alpha$ to $n^2(n-1)$.
In particular, pure Maxwell theory ($n=1$) does not allow a gauging of its duality symmetry.
This is in accordance with the explicit recent results of \cite{Deser:2010it}. 
For $n=2$, the linear constraints allow
for 4 independent parameters, however, it may be shown that the bilinear constraints (\ref{quadcon}) in this case do not admit any real solution. Non-trivial gaugings exist only for $n>2$ Maxwell fields.
For $n=3$ one finds that the non-trivial solution of the consistency constraints is unique up to conjugation 
and corresponds to gauging an $SO(3)$ subgroup of the global $U(3)$. This is a special case of the 
general construction given in the following.

Let us consider pure Yang-Mills theory with compact semi-simple 
gauge group $G$ and $n\equiv {\rm dim}\,G$
gauge fields $A_\mu{}^A$ with the standard embedding $G\subset SO(n)\subset U(n)$\,.
We denote by $f_{AB}{}^C$ and $\eta_{AB}$ the structure constants and the Cartan-Killing form 
of $G$, respectively. The duality covariant formulation of the theory will
employ $2n$ vector fields $A_\mu{}^M = (A_\mu{}^A, C_{\mu\,A})$, where for clarity 
we denote the dual vectors by $C_{\mu\,A}$. The embedding tensor is chosen such
that the generalized structure constants (\ref{gsc}) take the form
\bea
X_{MN}{}^K &\equiv& \left\{
\begin{array}{rcl}
X_{AB}{}^C &=&-f_{AB}{}^C\\
X_A{}^B{}_C &=& \phantom{-}f_{AC}{}^B{}
\end{array}\right.
\;,
\label{embYM}
\eea
with all other components vanishing. 
It is straightforward to verify that they satisfy the symplectic embedding (\ref{symplectic}),
and the linear constraint (\ref{linear}), while the bilinear constraints (\ref{quadcon}),
reduce to the standard Jacobi identities for the structure constants $f_{AB}{}^C$\,.
Moreover, this choice of the embedding tensor explicitly breaks the electric/magnetic duality $U(1)$
(even though the gauge group commutes with it).
The constant matrix ${\cal M}_{MN}$ is parametrized by
the Cartan-Killing form $\eta_{AB}$ and its inverse $\eta^{AB}$
\bea
{\cal M}_{MN} &=& 
\left(
\begin{array}{cc}
\eta_{AB}&0 \\
0&\eta^{AB} 
\end{array}\right)
\;.
\eea
The non-abelian twisted self-duality equations (\ref{eomNA1}) in this case take the form
\bea
{\cal F}_{\mu\nu}{}^A &=& -\ft12 \varepsilon_{\mu\nu\rho\sigma}\,\eta^{AB} {\cal G}^{\rho\sigma}{}_B
\;,
\label{FG}
\eea
with the field strengths
\bea
{\cal F}_{\mu\nu}{}^A &\equiv&  2\partial_{[\mu}A_{\nu]}{}^A -g f_{BC}{}^A \, A_\mu{}^B A_\nu{}^C 
\;,
\nonumber\\
{\cal G}_{\mu\nu}{}^A &=& 2D_{[\mu}C_{\nu]}{}^A + g{B}_{\mu\nu}{}^A
\;,
\eea
where we have lowered and raised indices $A, B, \dots$, with $\eta_{AB}$ and its inverse, respectively,
and redefined the two-forms according to 
${B}_{\mu\nu\,A}\equiv \frac12\Theta_A{}^\alpha B_{\mu\nu\,\alpha}+f_{ABC}{}\, A_{[\mu}{}^B C_{\nu]}{}^C$
with respect to the previous notation.
Gauge transformations (\ref{gaugeAB}) are given by
\bea
\delta A_\mu{}^A &=& D_\mu \Lambda^A\;,
\qquad
\delta C_\mu{}^A ~=~ D_\mu \tilde{\Lambda}^A -g \Lambda_\mu{}^A
\;,\nonumber\\[1ex]
\delta {B}_{\mu\nu}{}^A &=& 2 D_{[\mu} \Lambda_{\nu]}{}^A
+g f_{BC}{}^A\left(\Lambda^B {\cal G}_{\mu\nu}{}^C-\tilde{\Lambda}^B {\cal F}_{\mu\nu}{}^C\right) 
+ 2g f_{BC}{}^A C_{[\mu}{}^B \delta A_{\nu]}{}^C
\;,
\label{gtYM}
\eea
with independent parameters $\Lambda^A$, $\tilde{\Lambda}^A$ and $\Lambda_\mu{}^A$\,.
This illustrates explicitly how the two-form potential $B_{\mu\nu}{}^A$ together with its gauge invariance
cures the inconsistencies of a gauge theory with vector fields $C_\mu{}^A$ that do
not participate in the gauging but are charged under the gauge group.\footnote{
The explicit form of (\ref{gtYM}) shows that the construction in this example becomes somewhat degenerate: 
the gauge transformations $\tilde\Lambda^A$ of the dual vector fields $C_\mu{}^A$ 
may be entirely absorbed into a redefinition of the tensor gauge parameter $\Lambda_\mu{}^A$ .}
The non-abelian field strength (\ref{H3}) of the two-forms 
takes the form
\begin{eqnarray}
 {\cal H}_{\mu\nu\rho\,A}&=&3  D_{[\mu}{B}_{\nu\rho]\,A}
-3gf_{ABC} \, {\cal F}_{[\mu\nu}{}^B\, C_{\rho]}{}^C
\;,
\label{H3YM}
\end{eqnarray}
and it vanishes on-shell according to its equations of motion (\ref{eomNA2}).
It is straightforward to verify that (\ref{FG})
together with the first-order equation ${\cal H}_{\mu\nu\rho\,A}=0$ is equivalent to
the original Yang-Mills field equations for the vector field $A_\mu{}^A$\,.
The full Lagrangian (\ref{Lfull})  takes the explicit form
\bea
{\cal L}_{\rm covariant}
&=&
-\frac{1}{4}\,   {\cal F}_{jk}{}^A{\eta}_{AB} {\cal F}^{jk}{}^B
-\frac{1}{4}\,   {\cal G}_{jk}{}^A{\eta}_{AB} {\cal G}^{jk}{}^B
-\frac16 \,g \varepsilon^{ijk} \,
{\cal H}_{ijk\,A} \, A_0{}^A
\nonumber\\[.5ex]
&&
+\frac{1}{4}\,   \varepsilon^{ijk} \, \dot{C}_i{}^A {\eta}_{AB} {\cal F}_{jk}{}^B
-\frac{1}{4} \,  \varepsilon^{ijk} \,{\dot A}_{i}{}^A {\eta}_{AB} 
\left({\cal G}_{jk}{}^B+gB_{jk}{}^B\right)
\;,
\label{Lexp}
\eea
which gives the first-order Lagrangian for pure Yang-Mills theory,
describing the Yang-Mills vector fields $A_\mu{}^M$
together with their duals $C_\mu{}^A$. However, this Lagrangian 
no longer exhibits an analogue of the 
electric/magnetic $U(1)$ symmetry, since
the choice of the embedding tensor (\ref{embYM}) explicitly breaks this $U(1)$,
although the gauge group commutes with it.
This is in accordance with the explicit no-go results of~\cite{Deser:1976iy}.
The field equations derived from (\ref{Lexp})
are equivalent to the Yang-Mills equations $D^\mu {\cal F}_{\mu\nu}{}^A=0$,
while in the limit $g\rightarrow0$, the Lagrangian consistently reduces to the first-order 
duality invariant Lagrangian (\ref{actionHT}) describing a set of Maxwell equations.
On the other hand, it follows directly, that by integrating out the two-form potentials $B_{ij}{}^A$, 
also the magnetic vector fields $C_i{}^A$ disappear from (\ref{Lexp}),
and the Lagrangian reduces to the original second-order Yang-Mills form
\bea
\mathcal{L}_{\rm YM} &=&
- \frac14 \, \eta_{AB}\,\mathcal{F}_{\mu\nu}{}^{A} 
\mathcal{F}^{\mu\nu\,B}  \;,
\eea
which is manifestly four-dimensional covariant.

\section{Conclusions}
\label{sec:conclusions}

In this paper, we have constructed the general non-abelian deformations
of the first-order actions of~\cite{Henneaux:1988gg,Schwarz:1993vs},
whose field equations give rise to a non-abelian version of the
twisted self-duality equations (\ref{TSD}). Consistent gaugings are characterized by
the choice of a constant embedding tensor satisfying the set of algebraic 
constraints (\ref{linear}), (\ref{quadcon}). 
The resulting action (\ref{Lfull}) is formally invariant under the global duality group
$G_{\rm duality}$, if the embedding tensor itself transforms as a spurionic object 
under the duality group. The duality symmetry is broken when the embedding tensor
is set to a particular constant value.
Space-time covariance of the action is no longer manifest, but can be restored by
modifying the action of diffeomorphisms on the gauge fields by additional contributions
proportional to the equations of motion.
The construction relies on the introduction of additional couplings to higher-order $p$-forms
and a topological term. 

The set of algebraic consistency constraints for the embedding tensor coincides with
the one found for gaugings in the standard second-order formalism~\cite{deWit:2005ub}.
In other words, every gauging of the standard second-order action admits an analogous 
first-order action carrying all $2n$ vector fields. We call to mind that 
in the second-order formulation typically only a part of the global
duality group $G_{\rm duality}$ is realized as a symmetry of the ungauged action.
Nevertheless, more general subgroups of $G_{\rm duality}$ can be gauged upon the
introduction of magnetic vector fields. In the first-order formalism of this paper, all gaugings
are directly obtained as gauging of part of the off-shell symmetries $G_{\rm duality}$ of the 
first-order action (\ref{actionHT}).

Despite the equivalence of the resulting field equations, the first-order and second-order formulations
of gaugings remain complementary in several aspects: 
The second-order action~\cite{deWit:2005ub} carries 
as many two-form potentials as the gauging involves magnetic vector fields. In particular, there always exists 
a symplectic frame in which all vector fields involved in the gauging are electric and no two-forms appear in the action.
In contrast, the first-order action (\ref{Lfull}) inevitably carries a fixed number of two-form potentials 
equal to the dimension of the gauge group.
In the second-order formalism, space-time diffeomorphisms are realized in the standard way, while 
the canonical gauge transformations of the two-form potentials need to be modified by contributions
proportional to the field equations~\cite{deWit:2005ub}. Gauge transformations in the first-order formalism~(\ref{gaugeAB})
in contrast are of the canonical form whereas it is the action of space-time diffeomorphisms 
that is modified by on-shell vanishing contributions.
On the technical side of the construction, 
in the second-order formalism it is only the sum of kinetic and topological term of gauge fields
that is gauge invariant. In the first-order formalism these terms are separately gauge invariant,
but only a particular combination of them gives rise to a consistent set of field equations.

We emphasize that the construction of non-abelian gauge deformations of the first-order
action (\ref{actionHT}) inevitably requires the introduction of two-form potentials with
St\"uckelberg-type couplings to the vector fields~(\ref{defH}) and a $BF$-type coupling
in the topological term (\ref{Ltop}). It has recently been concluded in~\cite{Bunster:2010wv}
that in absence of two-form potentials the first-order action (\ref{actionHT})
does not admit {\em any} non-abelian deformation, see~\cite{Bekaert:2001wa} for earlier
no-go results. Specifically, the deformations studied in~\cite{Bunster:2010wv} are triggered
by antisymmetric 
structure constants  $C^K_{MN} \sim X_{[MN]}{}^K$, corresponding to the ansatz~(\ref{covder}),
(\ref{gsc}),
with the additional assumption that $X_{(MN)}{}^K\equiv0$. Indeed, in the general construction
of non-abelian deformations, it is the symmetric part of the generalized structure constants 
$X_{(MN)}{}^K$ that governs the introduction and additional 
couplings to two-form potentials~\cite{deWit:2005hv}. 
Imposing the absence of two-form potentials thus requires that $X_{(MN)}{}^K\equiv0$.
However, in four dimensions this additional constraint is in direct contradiction to the constraints on the 
embedding tensor, cf.\ footnote~\ref{fn:sym}, and implies that the $X_{MN}{}^K$
vanish identically, in
accordance with the analysis of~\cite{Bunster:2010wv}. 
Hence, there are no Yang-Mills-type deformations 
of the first-order action (\ref{actionHT}) 
in the absence of two-form potentials.
In contrast, we have shown that by proper inclusion of such higher-order $p$-forms, 
the first-order action can be consistently generalized to the non-abelian case.

The construction of the first-order action (\ref{Lfull}) for non-abelian gauge fields 
furthermore allows to revisit the recent discussion of possible gaugings of (subgroups of) 
electric/magnetic duality~\cite{Bunster:2010wv,Deser:2010it}.
The answer to the question which subgroups of the global duality group $G_{\rm duality}$
can be gauged, now turns out to be encoded in the set of algebraic consistency constraints
(\ref{linear}), (\ref{quadcon}) for the embedding tensor.
Indeed, we have shown in section~\ref{sec:YM} that for pure Maxwell theory ($n=1$, no scalar fields),
these constraints do not admit any solution, thus reproducing the no-go result of~\cite{Deser:2010it}.
In contrast, for a set of $n$ Maxwell fields, we find that any compact group of dimension~$n$
can be gauged via the standard embedding $G\subset SO(n)\subset U(n)$ into the global duality group.
The resulting Lagrangian explicitly breaks electric/magnetic duality, in accordance
with the analysis of~\cite{Deser:1976iy}.
In the presence of scalar fields, the global duality group of (\ref{actionHT}) is enlarged to some
non-compact subgroup of $Sp(2n,\mathbb{R})$ and the possibilities for admissible gauge groups
further increase, in particular due to the existence of different inequivalent symplectic frames. 
Many solutions to the corresponding consistency constraints have been found in the second-order 
formalism for extended gauged supergravities~\cite{Schon:2006kz,deWit:2007mt}.
E.g.\ the presented construction immediately gives rise to a deformation of the 
duality-invariant first-order action of~\cite{Hillmann:2009zf}, that describes the
maximal $SO(8)$ gauged supergravity of~\cite{deWit:1982ig}.

In this paper, we have restricted the discussion to vector fields in four space-time dimensions.
Further coupling to fermions and the realization of supersymmetry will proceed straightforwardly
along the lines of~\cite{Hillmann:2009zf}. 
Moreover, the embedding tensor formalism employed in this paper serves as a guide
as to how the presented construction can be straightforwardly
generalized to describe non-abelian deformations of arbitrary $p$-forms in $D$ space-time
dimensions.
The key structure underlying these constructions is the structure of non-abelian deformations
of the associated hierarchy of $p$-form 
tensor fields~\cite{deWit:2005hv,deWit:2008ta,Bergshoeff:2009ph}.
In analogy to the theories constructed in this paper, the deformations of first-order 
actions of $p$-forms will generically involve couplings to higher-order $\mbox{$(p+1)$}$-forms
and necessitate contributions from additional topological terms.
While in the example of Yang-Mills theory discussed in the previous section,
the introduction of additional two-form potentials may still occur as somewhat artificial,
in higher-dimensional supergravity theories 
typically the full hierarchy of anti-symmetric $p$-form fields is already present in 
the original theory. It appears rather natural that all of them should be involved in the 
new couplings of the first-order action.
Such a construction should culminate in a first-order description of the higher-dimensional
theories in which all $p$-forms are present, related to their duals by the proper
non-abelian extension of the actions of~\cite{Henneaux:1988gg,Schwarz:1993vs,Bunster:2011aw},
and linked to their adjacent forms by St\"uckelberg-type couplings analogous to (\ref{defH}).
This formulation may be particularly useful in the study of dimensional
reductions and the realization of the hidden symmetries 
$E_{10}$, $E_{11}$, \cite{West:2001as,Damour:2002cu}, extending the duality symmetries.

Another obvious avenue of generalization of the presented construction is the lift of
the non-abelian deformations to the non-polynomial Lagrangians with manifest 
space-time symmetry~\cite{Pasti:1995tn,Pasti:1996vs} from which the first-order
actions~\cite{Henneaux:1988gg,Schwarz:1993vs} are obtained after particular gauge fixing. 
Also the generalization to deformations of the first-order formulations based on a 
more general decomposition of space-time into $d+(D-d)$ dimensions~\cite{Chen:2010jgb} 
(generalizing the case $d=1$ discussed in this paper) should be achievable along similar lines.

Finally, a particularly interesting realm of applications of the first-order actions (\ref{actionHT})
and their deformations described in this paper, are those chiral theories which do not admit
the formulation of a standard second-order action.
In this context, we mention most notably the six-dimensional chiral tensor gauge theories 
and recent attempts to construct their non-abelian deformations
supposed to underlie the description of
multiple M5-brane dynamics~\cite{Perry:1996mk,Howe:1997vn,Pasti:2009xc,Lambert:2010wm,Ho:2011ni}.
The construction presented in this paper may provide a well-defined framework for the general
study of such interactions.

\vspace{8mm}
\noindent
{\bf Acknowledgement}\\
\noindent
This work is supported in part by the Agence Nationale de la Recherche (ANR).
I wish to thank M.\ Magro for helpful discussions.
\bigskip


\providecommand{\href}[2]{#2}\begingroup\raggedright\endgroup

\end{document}